\documentclass[%
 reprint,
superscriptaddress,
showpacs,
 amsmath,amssymb,
 aps,
]{revtex4-1}

\usepackage{graphicx}
\usepackage{dcolumn}
\usepackage{bm}
\usepackage{amsmath}
\usepackage{autobreak}
\usepackage{hyperref}
\usepackage[mathlines]{lineno}
\usepackage{natbib}
\usepackage{multirow}
\usepackage{makecell}
\usepackage{placeins}

\begin{document}

\preprint{APS/123-QED}

\title{Constraints on inelastic dark matter from the CDEX-1B experiment}

\author{Y.~F.~Liang}
\affiliation{Key Laboratory of Particle and Radiation Imaging (Ministry of Education) and Department of Engineering Physics, Tsinghua University, Beijing 100084}
\author{L.~T.~Yang}\altaffiliation [Corresponding author: ]{yanglt@mail.tsinghua.edu.cn}
\affiliation{Key Laboratory of Particle and Radiation Imaging (Ministry of Education) and Department of Engineering Physics, Tsinghua University, Beijing 100084}
\author{Q. Yue}\altaffiliation [Corresponding author: ]{yueq@mail.tsinghua.edu.cn}
\affiliation{Key Laboratory of Particle and Radiation Imaging (Ministry of Education) and Department of Engineering Physics, Tsinghua University, Beijing 100084}

\author{K.~J.~Kang}
\affiliation{Key Laboratory of Particle and Radiation Imaging (Ministry of Education) and Department of Engineering Physics, Tsinghua University, Beijing 100084}
\author{Y.~J.~Li}
\affiliation{Key Laboratory of Particle and Radiation Imaging (Ministry of Education) and Department of Engineering Physics, Tsinghua University, Beijing 100084}

\author{H.~P.~An}
\affiliation{Key Laboratory of Particle and Radiation Imaging (Ministry of Education) and Department of Engineering Physics, Tsinghua University, Beijing 100084}
\affiliation{Department of Physics, Tsinghua University, Beijing 100084}

\author{Greeshma~C.}
\altaffiliation{Participating as a member of TEXONO Collaboration}
\affiliation{Institute of Physics, Academia Sinica, Taipei 11529}

\author{J.~P.~Chang}
\affiliation{NUCTECH Company, Beijing 100084}
\author{H.~Chen}
\affiliation{Key Laboratory of Particle and Radiation Imaging (Ministry of Education) and Department of Engineering Physics, Tsinghua University, Beijing 100084}

\author{Y.~H.~Chen}
\affiliation{YaLong River Hydropower Development Company, Chengdu 610051}
\author{J.~P.~Cheng}
\affiliation{Key Laboratory of Particle and Radiation Imaging (Ministry of Education) and Department of Engineering Physics, Tsinghua University, Beijing 100084}
\affiliation{School of Physics and Astronomy, Beijing Normal University, Beijing 100875}
\author{J.~Y.~Cui}
\affiliation{Key Laboratory of Particle and Radiation Imaging (Ministry of Education) and Department of Engineering Physics, Tsinghua University, Beijing 100084}
\author{W.~H.~Dai}
\affiliation{Key Laboratory of Particle and Radiation Imaging (Ministry of Education) and Department of Engineering Physics, Tsinghua University, Beijing 100084}
\author{Z.~Deng}
\affiliation{Key Laboratory of Particle and Radiation Imaging (Ministry of Education) and Department of Engineering Physics, Tsinghua University, Beijing 100084}
\author{Y.~X.~Dong}
\affiliation{Key Laboratory of Particle and Radiation Imaging (Ministry of Education) and Department of Engineering Physics, Tsinghua University, Beijing 100084}
\author{C.~H.~Fang}
\affiliation{College of Physics, Sichuan University, Chengdu 610065}

\author{H.~Gong}
\affiliation{Key Laboratory of Particle and Radiation Imaging (Ministry of Education) and Department of Engineering Physics, Tsinghua University, Beijing 100084}
\author{Q.~J.~Guo}
\affiliation{School of Physics, Peking University, Beijing 100871}
\author{T.~Guo}
\affiliation{Key Laboratory of Particle and Radiation Imaging (Ministry of Education) and Department of Engineering Physics, Tsinghua University, Beijing 100084}
\author{X.~Y.~Guo}
\affiliation{YaLong River Hydropower Development Company, Chengdu 610051}
\author{L.~He}
\affiliation{NUCTECH Company, Beijing 100084}
\author{J.~R.~He}
\affiliation{YaLong River Hydropower Development Company, Chengdu 610051}

\author{H.~X.~Huang}
\affiliation{Department of Nuclear Physics, China Institute of Atomic Energy, Beijing 102413}
\author{T.~C.~Huang}
\affiliation{Sino-French Institute of Nuclear and Technology, Sun Yat-sen University, Zhuhai 519082}

\author{S.~Karmakar}
\altaffiliation{Participating as a member of TEXONO Collaboration}
\affiliation{Institute of Physics, Academia Sinica, Taipei 11529}

\author{Y.~S.~Lan}
\affiliation{Key Laboratory of Particle and Radiation Imaging (Ministry of Education) and Department of Engineering Physics, Tsinghua University, Beijing 100084}
\author{H.~B.~Li}
\altaffiliation{Participating as a member of TEXONO Collaboration}
\affiliation{Institute of Physics, Academia Sinica, Taipei 11529}
\author{H.~Y.~Li}
\affiliation{College of Physics, Sichuan University, Chengdu 610065}
\author{J.~M.~Li}
\affiliation{Key Laboratory of Particle and Radiation Imaging (Ministry of Education) and Department of Engineering Physics, Tsinghua University, Beijing 100084}
\author{J.~Li}
\affiliation{Key Laboratory of Particle and Radiation Imaging (Ministry of Education) and Department of Engineering Physics, Tsinghua University, Beijing 100084}
\author{M.~C.~Li}
\affiliation{YaLong River Hydropower Development Company, Chengdu 610051}
\author{Q.~Y.~Li}
\affiliation{College of Physics, Sichuan University, Chengdu 610065}
\author{R.~M.~J.~Li}
\affiliation{College of Physics, Sichuan University, Chengdu 610065}
\author{X.~Q.~Li}
\affiliation{School of Physics, Nankai University, Tianjin 300071}
\author{Y.~L.~Li}
\affiliation{Key Laboratory of Particle and Radiation Imaging (Ministry of Education) and Department of Engineering Physics, Tsinghua University, Beijing 100084}

\author{B.~Liao}
\affiliation{School of Physics and Astronomy, Beijing Normal University, Beijing 100875}
\author{F.~K.~Lin}
\altaffiliation{Participating as a member of TEXONO Collaboration}
\affiliation{Institute of Physics, Academia Sinica, Taipei 11529}
\author{S.~T.~Lin}
\affiliation{College of Physics, Sichuan University, Chengdu 610065}
\author{J.~X.~Liu}
\affiliation{Key Laboratory of Particle and Radiation Imaging (Ministry of Education) and Department of Engineering Physics, Tsinghua University, Beijing 100084}
\author{R.~Z.~Liu}
\affiliation{Key Laboratory of Particle and Radiation Imaging (Ministry of Education) and Department of Engineering Physics, Tsinghua University, Beijing 100084}
\author{S.~K.~Liu}
\affiliation{College of Physics, Sichuan University, Chengdu 610065}
\author{Y.~D.~Liu}
\affiliation{School of Physics and Astronomy, Beijing Normal University, Beijing 100875}
\author{Y.~Liu}
\affiliation{College of Physics, Sichuan University, Chengdu 610065}
\author{Y.~Y.~Liu}
\affiliation{School of Physics and Astronomy, Beijing Normal University, Beijing 100875}
\author{H.~Ma}
\affiliation{Key Laboratory of Particle and Radiation Imaging (Ministry of Education) and Department of Engineering Physics, Tsinghua University, Beijing 100084}
\author{Y.~C.~Mao}
\affiliation{School of Physics, Peking University, Beijing 100871}
\author{A.~Mureed}
\affiliation{College of Physics, Sichuan University, Chengdu 610065}
\author{H.~Pan}
\affiliation{NUCTECH Company, Beijing 100084}
\author{N.~C.~Qi}
\affiliation{YaLong River Hydropower Development Company, Chengdu 610051}
\author{J.~Ren}
\affiliation{Department of Nuclear Physics, China Institute of Atomic Energy, Beijing 102413}
\author{X.~C.~Ruan}
\affiliation{Department of Nuclear Physics, China Institute of Atomic Energy, Beijing 102413}
\author{M.~B.~Shen}
\affiliation{YaLong River Hydropower Development Company, Chengdu 610051}
\author{H.~Y.~Shi}
\affiliation{College of Physics, Sichuan University, Chengdu 610065}
\author{M.~K.~Singh}
\altaffiliation{Participating as a member of TEXONO Collaboration}
\affiliation{Institute of Physics, Academia Sinica, Taipei 11529}
\affiliation{Department of Physics, Banaras Hindu University, Varanasi 221005}
\author{T.~X.~Sun}
\affiliation{School of Physics and Astronomy, Beijing Normal University, Beijing 100875}
\author{W.~L.~Sun}
\affiliation{YaLong River Hydropower Development Company, Chengdu 610051}
\author{C.~J.~Tang}
\affiliation{College of Physics, Sichuan University, Chengdu 610065}
\author{Y.~Tian}
\affiliation{Key Laboratory of Particle and Radiation Imaging (Ministry of Education) and Department of Engineering Physics, Tsinghua University, Beijing 100084}
\author{H.~F.~Wan}
\affiliation{Key Laboratory of Particle and Radiation Imaging (Ministry of Education) and Department of Engineering Physics, Tsinghua University, Beijing 100084}
\author{G.~F.~Wang}
\affiliation{School of Physics and Astronomy, Beijing Normal University, Beijing 100875}
\author{J.~Z.~Wang}
\affiliation{Key Laboratory of Particle and Radiation Imaging (Ministry of Education) and Department of Engineering Physics, Tsinghua University, Beijing 100084}
\author{L.~Wang}
\affiliation{School of Physics and Astronomy, Beijing Normal University, Beijing 100875}
\author{Q.~Wang}
\affiliation{College of Physics, Sichuan University, Chengdu 610065}
\author{Q.~Wang}
\affiliation{Key Laboratory of Particle and Radiation Imaging (Ministry of Education) and Department of Engineering Physics, Tsinghua University, Beijing 100084}
\affiliation{Department of Physics, Tsinghua University, Beijing 100084}
\author{Y.~F.~Wang}
\affiliation{Key Laboratory of Particle and Radiation Imaging (Ministry of Education) and Department of Engineering Physics, Tsinghua University, Beijing 100084}
\author{Y.~X.~Wang}
\affiliation{School of Physics, Peking University, Beijing 100871}
\author{H.~T.~Wong}
\altaffiliation{Participating as a member of TEXONO Collaboration}
\affiliation{Institute of Physics, Academia Sinica, Taipei 11529}

\author{Y.~C.~Wu}
\affiliation{Key Laboratory of Particle and Radiation Imaging (Ministry of Education) and Department of Engineering Physics, Tsinghua University, Beijing 100084}
\author{H.~Y.~Xing}
\affiliation{College of Physics, Sichuan University, Chengdu 610065}
\author{K.~Z.~Xiong}
\affiliation{YaLong River Hydropower Development Company, Chengdu 610051}
\author{R.~Xu}
\affiliation{Key Laboratory of Particle and Radiation Imaging (Ministry of Education) and Department of Engineering Physics, Tsinghua University, Beijing 100084}
\author{Y.~Xu}
\affiliation{School of Physics, Nankai University, Tianjin 300071}
\author{T.~Xue}
\affiliation{Key Laboratory of Particle and Radiation Imaging (Ministry of Education) and Department of Engineering Physics, Tsinghua University, Beijing 100084}
\author{Y.~L.~Yan}
\affiliation{College of Physics, Sichuan University, Chengdu 610065}
\author{N.~Yi}
\affiliation{Key Laboratory of Particle and Radiation Imaging (Ministry of Education) and Department of Engineering Physics, Tsinghua University, Beijing 100084}
\author{C.~X.~Yu}
\affiliation{School of Physics, Nankai University, Tianjin 300071}
\author{H.~J.~Yu}
\affiliation{NUCTECH Company, Beijing 100084}
\author{X.~Yu}
\affiliation{Key Laboratory of Particle and Radiation Imaging (Ministry of Education) and Department of Engineering Physics, Tsinghua University, Beijing 100084}
\author{M.~Zeng}
\affiliation{Key Laboratory of Particle and Radiation Imaging (Ministry of Education) and Department of Engineering Physics, Tsinghua University, Beijing 100084}
\author{Z.~Zeng}
\affiliation{Key Laboratory of Particle and Radiation Imaging (Ministry of Education) and Department of Engineering Physics, Tsinghua University, Beijing 100084}

\author{F.~S.~Zhang}
\affiliation{School of Physics and Astronomy, Beijing Normal University, Beijing 100875}

\author{P.~Zhang}
\affiliation{Key Laboratory of Particle and Radiation Imaging (Ministry of Education) and Department of Engineering Physics, Tsinghua University, Beijing 100084}

\author{P.~Zhang}
\affiliation{YaLong River Hydropower Development Company, Chengdu 610051}

\author{Z.~Y.~Zhang}
\affiliation{Key Laboratory of Particle and Radiation Imaging (Ministry of Education) and Department of Engineering Physics, Tsinghua University, Beijing 100084}

\author{M.~G.~Zhao}
\affiliation{School of Physics, Nankai University, Tianjin 300071}

\author{J.~F.~Zhou}
\affiliation{YaLong River Hydropower Development Company, Chengdu 610051}
\author{Z.~Y.~Zhou}
\affiliation{Department of Nuclear Physics, China Institute of Atomic Energy, Beijing 102413}
\author{J.~J.~Zhu}
\affiliation{College of Physics, Sichuan University, Chengdu 610065}

\collaboration{CDEX Collaboration}
\noaffiliation

\date{\today}

\begin{abstract}
We present limits on spin-independent inelastic weakly interacting massive particles (WIMP)-nucleus scattering using the 737.1 kg$\cdot$day dataset from the CDEX-1B experiment. Expected nuclear recoil spectra for various inelastic WIMP masses $m_\chi$ and mass splittings $\delta$ are calculated under the standard halo model. An accurate background model of CDEX-1B is constructed by simulating all major background sources. The model parameters are then determined through maximum likelihood estimation and Markov chain Monte Carlo fitting. The resulting 90\% confidence level upper limits on the WIMP-nucleon cross section $\sigma_{\mathrm{n}}$ exclude certain DAMA/LIBRA allowed regions: the $\chi^2 < 4$ regions for $\delta < 30$ keV at $m_\chi = 250$ GeV and the $\chi^2 < 9$ region for $\delta < 50$ keV at $m_\chi = 500$ GeV. The method is applicable to other inelastic dark matter scenarios, and the upcoming CDEX-50 experiment is expected to improve sensitivity by four orders of magnitude.
\end{abstract}
\maketitle

\section{\label{sec1}Introduction}
A substantial body of cosmological and astronomical evidence demonstrates that dark matter constitutes a fundamental component of the Universe~\cite{pdg2024,bertone_particle_2005}. The investigation of dark matter is one of the most critical challenges in modern physics. Weakly interacting massive particles (WIMPs) are the most popular dark matter candidates~\cite{pdg2024}. Numerous experiments have been dedicated to the direct detection of WIMPs, such as XENON~\cite{XENONnT}, PandaX~\cite{PandaX-4T}, LUX-ZEPLIN~\cite{LZ}, DarkSide~\cite{darkside}, SuperCDMS~\cite{cdmslite}, EDELWEISS~\cite{EDELWEISS}, CRESST~\cite{cresst}, DAMA/LIBRA~\cite{DAMA2010}, and CDEX~\cite{cdex1,cdex0,cdex12014,cdex12016,cdex1b2018,cdex102018,cdex10_tech,cdex1b_am,cdex10_eft,CDEX50pre,cdexmidgal}. However, WIMPs have not been detected to date. One possible reason is that elastic scattering between WIMPs and nuclei is heavily suppressed. Thus, the inelastic dark matter (iDM) scenario was proposed~\cite{SI_iDM}, in which WIMPs scatter with nuclei inelastically.

In inelastic WIMP-nucleus scattering, the WIMP is excited to a higher-energy state ($\chi^*$), characterized by a mass splitting $\delta$ from its ground state ($\chi$)~\cite{SI_iDM}. For a given nuclear recoil energy $E_{\mathrm{nr}}$, there exists a minimal required relative velocity of WIMPs,
\begin{equation} \label{eq:vmin}
  \begin{aligned}
    v_{\min }=\frac{1}{\sqrt{2 E_{\mathrm{nr}} m_{\mathrm{N}}}}\left(\frac{E_{\mathrm{nr}} m_{\mathrm{N}}}{\mu}+\delta\right),
  \end{aligned}
  \end{equation}
where $m_{\mathrm{N}}$ is the mass of the target nucleus, and $\mu$ is the reduced mass of the system. When $\delta = 0$, the scattering becomes elastic. If $E_{\mathrm{nr}}$ is too large or too small, $v_{\min}$ will exceed the maximum velocity of WIMPs that can reach the laboratory, thereby preventing such events from occurring. The maximum and minimum possible values of $E_{\mathrm{nr}}$ for a certain maximum WIMP velocity $v_{\max}$ are given by
\begin{equation} \label{eq:Enr}
  \begin{aligned}
    E_{\mathrm{nr,\ max/min}} = \frac{\mu^2 v_{\max}^2}{2m_{\mathrm{N}}} \left( 1 \pm \sqrt{1 - \frac{2\delta}{\mu v_{\max}^2}} \right)^2.
  \end{aligned}
  \end{equation}
Due to the limits on $E_{\mathrm{nr}}$, iDM exhibits greater sensitivity to the velocity distribution of dark matter than elastic dark matter.

Moreover, for a given $\delta$, a minimum WIMP velocity is required for the occurrence of the scattering,
\begin{equation} \label{eq:vgm}
  \begin{aligned}
    v_{\min}^{*} = \sqrt{\frac{2\delta}{\mu}},
  \end{aligned}
  \end{equation}
which indicates that if $\delta$ exceeds a certain threshold, the inelastic scattering will not occur in laboratory experiments due to excessive $v_{\mathrm{min}}^{*} $. 

Analogously to the case of elastic WIMPs, the interaction between inelastic WIMPs and nuclei at initial state ($A_i$) could be either spin independent (SI)~\cite{SI_iDM,U1B-L} or spin dependent~\cite{MiDM,EFTiDM}. This paper focuses exclusively on the SI scenario. In SI scenario, for $\delta < 1.022$ MeV, the deexcitation of excited-state WIMPs is considered to solely release neutrino-antineutrino pairs~\cite{U1B-L}, which are undetectable in conventional dark matter detectors. Consequently, the nuclear recoil energy is the only observable signature. According to Eq.~(\ref{eq:vgm}), inelastic scattering of Galactic WIMPs with $\delta \geq 1.022$ MeV is kinematically forbidden in most experimental setups due to the Galactic escape velocity~\cite{SHM} constraint. That is, the physics channel for this analysis is
\begin{equation} \label{eq:chi_A}
  \begin{gathered}
    \chi + A_i \;\rightarrow\; \chi^{*} + A_f, \\
    \chi^{*} \;\rightarrow\; \chi + \nu + \bar{\nu},
  \end{gathered}
  \end{equation}
where the possible nuclear recoil energy $E_{\mathrm{nr}}$ of the nuclei at their final state ($A_f$) are the measureables.

In this study, we place constraints on the inelastic WIMP-nucleon SI interactions with 737.1 kg$\cdot$day of data from the CDEX-1B experiment~\cite{cdex1b2018} at the China Jinping Underground Laboratory (CJPL)~\cite{cjpl}. The devised methodology can be adopted to study a class of iDM models, such as magnetic inelastic dark matter~\cite{MiDM}, effective field theory inelastic dark matter~\cite{EFTiDM}, inelastic Dirac dark matter~\cite{iDDM}, and inelastic boosted dark matter~\cite{iBDM1,iBDM2}.

\section{\label{sec2} Expected iDM spectra}
The differential nuclear recoil spectrum of inelastic WIMP-nucleus scattering is given by
\begin{equation} \label{eq:dRdEnr}
  \begin{aligned}
    \frac{\mathrm{d}R}{\mathrm{d}E_{\mathrm{nr}}} = \frac{\rho N_{\mathrm{T}}}{m_{\chi}} \int_{v_{\min}}^{v_{\max}} \frac{\mathrm{d}\sigma}{\mathrm{d}E_{\mathrm{nr}}}  v f(\vec{v}, t)  \mathrm{d}^3 v,
  \end{aligned}
  \end{equation}
where $N_{\mathrm{T}}$ is the number of target nuclei per unit effective mass of the detector, $\rho$ is the local density of WIMPs, $m_\chi$ is the mass of WIMPs, $d\sigma/dE$ is the differential cross section of the inelastic scattering, $f(v)$ is the velocity distribution of WIMPs in the rest frame of the Earth, the lower limit $v_{\min}$ is formulated by Eq.~(\ref{eq:vmin}), and the upper limit $v_{\max}$ is determined by the escape velocity of WIMPs in the Galaxy and the velocity of the Earth~\cite{SI_iDM,PandaX-II}. Assuming that the inelastic scattering is spin and energy independent, the differential cross section can be expressed as
\begin{equation} \label{eq:dSigmadE}
  \begin{aligned}
    \frac{d\sigma}{dE_{\mathrm{nr}}} = \frac{m_{\mathrm{N}} \sigma_{\mathrm{n}} }{2 \mu^2 v^2} \cdot (Z \cdot f^{\mathrm{p}} + (A-Z) \cdot f^{\mathrm{n}})^2 F^2_{\mathrm{SI}}(E_{\mathrm{nr}}),
  \end{aligned}
  \end{equation}
where $\sigma_{\mathrm{n}}$ is the SI WIMP-nucleon cross section, $\mu$ is the reduced mass of the WIMP-nucleon system, $Z$ is the atomic number of the nucleus, $A$ is the mass number of the nucleus, and $f^{\mathrm{p, n}}$ are the effective WIMP couplings to the proton and neutron, $F_{\mathrm{SI}}$ is the SI nuclear form factor~\cite{SI_iDM,PandaX-II}. 

High-purity germanium (HPGe) detectors~\cite{HPGe1,HPGe2}, owing to their good energy resolutions and ultralow energy thresholds, have been applied in dark matter direct detection by CDEX~\cite{cdex1,cdex0,cdex12014,cdex12016,cdex1b2018,cdex102018,cdex10_tech,cdex1b_am,cdex10_eft,cdexmidgal,CDEX50pre,cdex_darkphoton,cdex_crdm,cdex_bdm,cdex_blackhole,cdex_blackhole_e,cdex_exotic,cdex_fermionic,cdex_solarreflect,cdex_axion,CDEX_DM_e,cdex_lightmediator}. When WIMPs scatter with germanium (Ge) nuclei in the HPGe detector, a portion of the nuclear recoil energy will be converted into detectable ionization energy. This converted energy, called the electron-equivalent energy, is given by $E_{\mathrm{det}} = Q_{\mathrm{nr}}(E_{\mathrm{nr}}) \cdot E_{\mathrm{nr}}$, where $Q_{\mathrm{nr}}$ denotes the quenching factor~\cite{HPGe2}. The differential electron-equivalent spectrum of inelastic WIMP-nucleus scattering in the HPGe detector is given by
\begin{equation} \label{eq:dRdEdet}
  \begin{aligned}
    \frac{dR}{dE_{\mathrm{det}}} = \left( \frac{dQ^{-1}_{\mathrm{nr}}(E_{\mathrm{det}})}{dE_{\mathrm{det}}} E_{\mathrm{det}} + Q^{-1}_{\mathrm{nr}}(E_{\mathrm{det}}) \right) \frac{dR}{dE_{\mathrm{nr}}},
  \end{aligned}
  \end{equation}
where $Q_{\mathrm{nr}}^{-1}$ is the inverse function of the quenching factor function.

We adopt the standard halo model~\cite{SHM}. Accordingly, the WIMP density $\rho$ is fixed at 0.3 GeV/$(c^2\mathrm{cm}^3)$, the WIMP velocity distribution follows a Maxwellian profile with most probable velocity $v_0 = 238$ km/s, and the Galactic escape velocity is set to 544 km/s. For the nuclear form factor, the Helm parametrization~\cite{formfactor1, formfactor2} is employed. The quenching factor $Q_{\mathrm{nr}}$ is obtained from the average values provided by the \texttt{TRIM} simulation package~\cite{TRIM}, and these values differ from the Lindhard model~\cite{LindhardModel} prediction with $\kappa  = 0.162$ (as adopted in Ref.~\cite{Lindhard162}) by less than 12\% at the energies relevant to this work. For WIMPs with $m_\chi = 100$ GeV, the resulting impact on the derived limits for inelastic WIMP-nucleon SI interactions remains within 30\%.

\begin{figure}[htbp]
  \includegraphics[width=\linewidth]{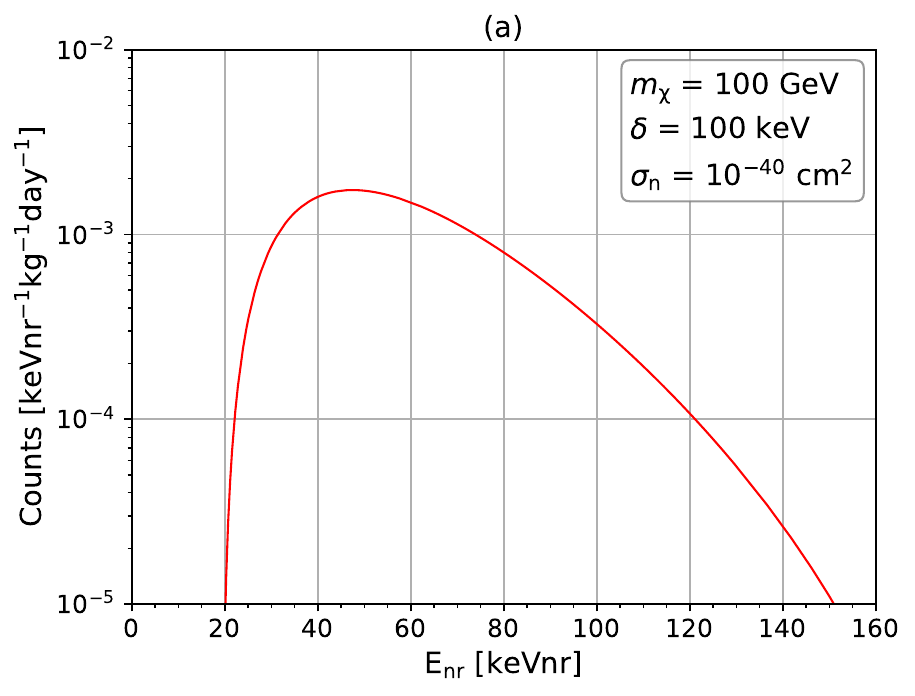}
  \includegraphics[width=\linewidth]{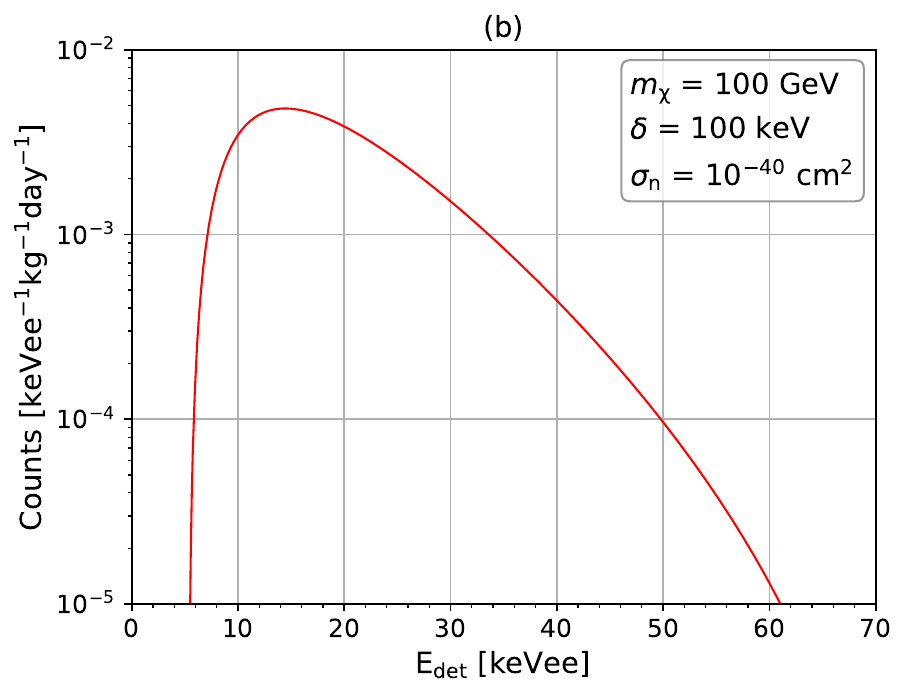}
  \caption{
    Expected (a) nuclear recoil and (b) electron-equivalent spectra of inelastic WIMP-nucleus scattering in the HPGe detector with $m_\chi = 100$ GeV, $\delta = 100$ keV, and $\sigma_{\mathrm{n}} = 10^{-40}\ \mathrm{cm}^2$. The energy resolution of the CDEX-1B detector~\cite{cdex1b2018} is applied to the electron-equivalent spectrum.
  }
  \label{fig::nr_det_exp}
  \end{figure}

Figure~\ref{fig::nr_det_exp} shows the expected nuclear recoil and electron-equivalent spectra of inelastic WIMP-nucleus scattering in the HPGe detector, assuming $m_\chi = 100$ GeV, $\delta = 100$ keV, and $\sigma_{\mathrm{n}} = 10^{-40}\ \mathrm{cm}^2$. In the electron-equivalent spectrum, we incorporate the energy resolution of the HPGe detector applied in the CDEX-1B experiment~\cite{cdex1b2018}.

\begin{figure}[htbp]
  \includegraphics[width=\linewidth]{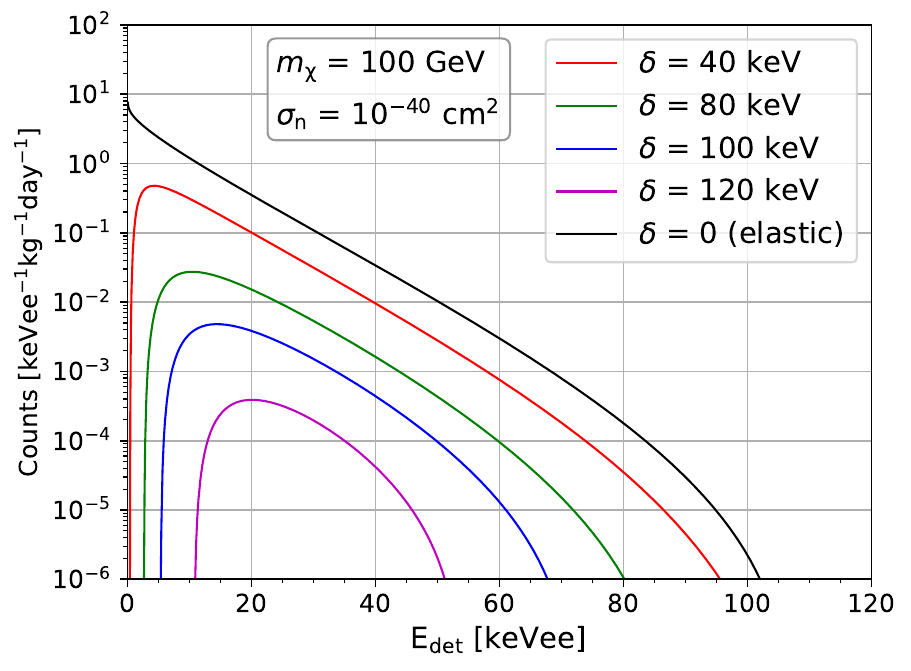}
  \caption{
  The expected spectra of inelastic WIMP-nucleus scattering in the HPGe detector for $\delta$ = 40, 80, 100, and 120  keV with $m_\chi = 100$ GeV and $\sigma_{\mathrm{n}} = 10^{-40}\ \mathrm{cm}^2$, compared with the expected spectrum of the elastic scattering.
  }
  \label{fig::det_exp_delta}
  \end{figure}

  Figure~\ref{fig::det_exp_delta} shows the expected spectra of inelastic WIMP-nucleus scattering in the HPGe detector for different $\delta$ values, with fixed $m_\chi$ and $\sigma_{\mathrm{n}}$. It also shows the expected spectrum for elastic scattering for comparison. From Fig.~\ref{fig::det_exp_delta}, it can be observed that the distinction between inelastic and elastic scattering on the expected nuclear recoil spectrum primarily manifests as a heavy suppression in the low-energy region of the inelastic scattering spectrum. In addition, Fig.~\ref{fig::det_exp_delta} demonstrates that with increasing $\delta$, the event rate of the entire nuclear recoil spectrum decreases, and the suppressed region expands to higher energies.

\begin{figure}[htbp]
  \includegraphics[width=\linewidth]{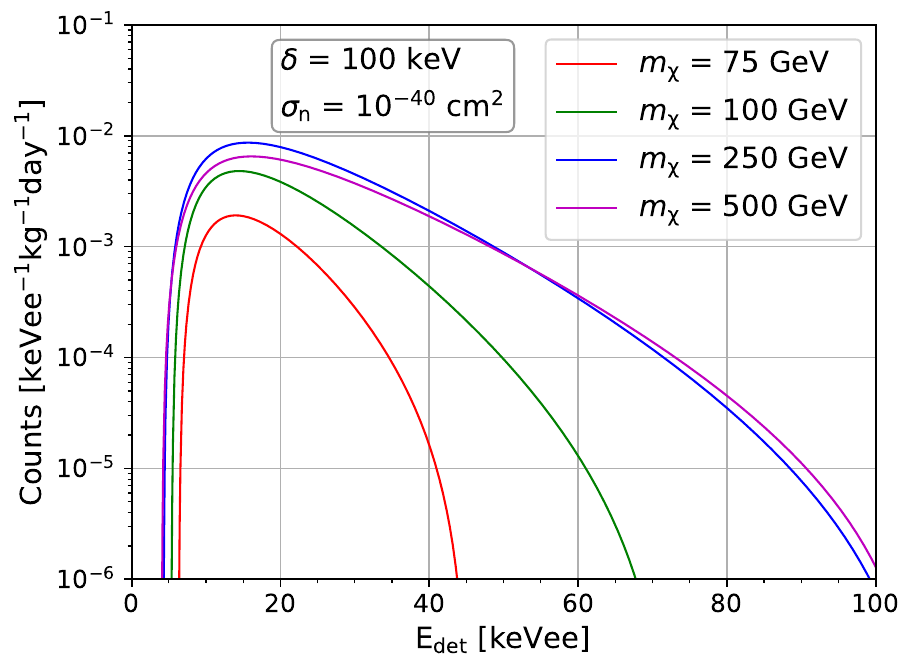}
  \caption{
    Expected spectra of the inelastic scattering in the HPGe detector for $m_\chi$ = 75, 100, 250, and 500 GeV with $\delta = 100$ keV and $\sigma_{\mathrm{n}} = 10^{-40}\ \mathrm{cm}^2$.
  }
  \label{fig::det_exp_Mx}
  \end{figure}

  Figure~\ref{fig::det_exp_Mx} shows the expected inelastic scattering spectra in the HPGe detector for different $m_\chi$ values, with fixed $\delta$ and $\sigma_{\mathrm{n}}$. As illustrated in Fig.~\ref{fig::det_exp_Mx}, the nuclear recoil spectrum falls off less sharply as $m_\chi$ increases.

\section{\label{sec3} CDEX-1B experiment}
The CDEX experiment utilizes HPGe detectors for direct dark matter detection~\cite{cdex1,cdex0,cdex12014,cdex12016,cdex1b2018,cdex102018,cdex10_tech,cdex1b_am,cdex10_eft,cdexmidgal,CDEX50pre,cdex_darkphoton,cdex_crdm,cdex_bdm,cdex_blackhole,cdex_blackhole_e,cdex_exotic,cdex_fermionic,cdex_solarreflect,cdex_axion,CDEX_DM_e,cdex_lightmediator} at CJPL~\cite{cjpl}, located beneath 2400 meters of rock overburden. The CDEX-1B experiment~\cite{cdex1b2018} runs one p-type point contact HPGe detector with a 1008 g target mass (fiducial mass of 939 g, after corrections due to a 0.88 $\pm$ 0.12 mm surface layer~\cite{MJL:deadlayer}) cooled by a cold finger. Additionally, a NaI(Tl) anti-Compton detector is employed to veto background events. With this setup, the experiment demonstrated excellent time stability, low background level, and good energy resolution~\cite{cdex1b2018,cdex1b_am}.

\begin{figure*}[htbp]
  \includegraphics[width=\linewidth]{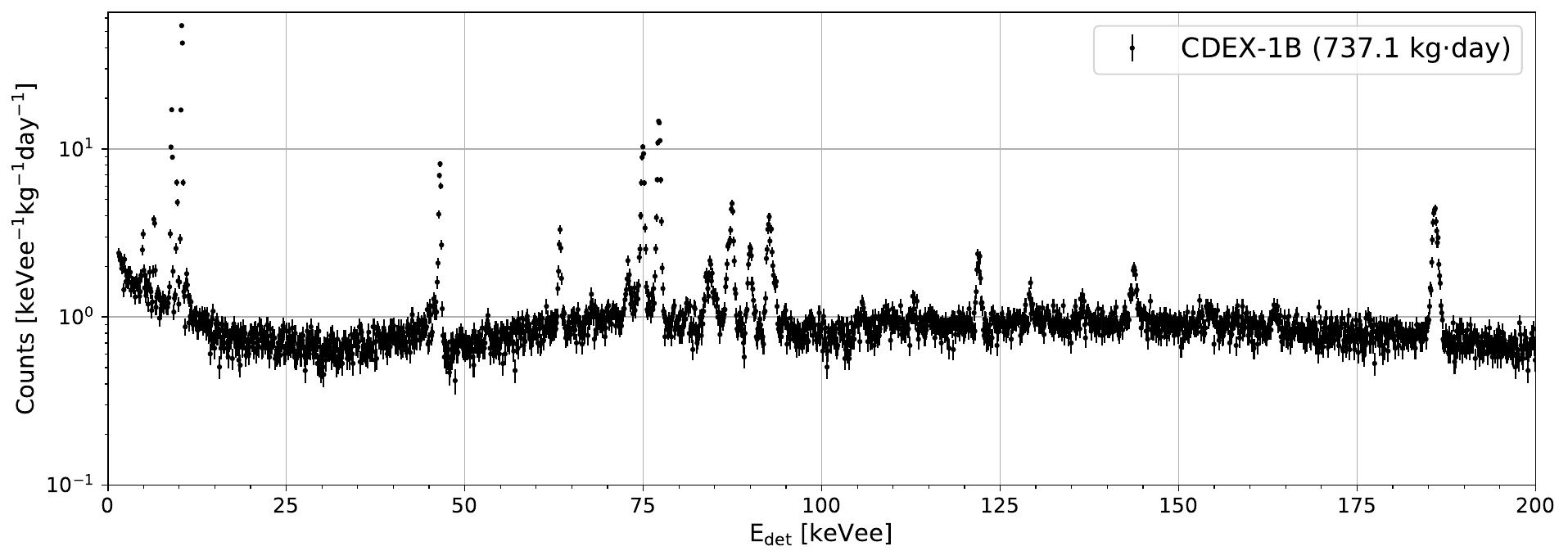}
  \caption{
    Background spectrum with error bars based on the 737.1 kg$\cdot$day dataset of the CDEX-1B experiment~\cite{cdex1b2018}. The spectrum is processed using anti-Compton veto and bulk-surface event discrimination~\cite{cdex1b2018}. The bin width is 100 eVee and the energy range is 1.5--200 keVee. 
  }
  \label{fig::C1B_bkg}
  \end{figure*}

The background spectrum of the CDEX-1B experiment, based on the 737.1 kg$\cdot$day dataset collected from March 2014 to July 2017~\cite{cdex1b2018}, is shown in Fig.~\ref{fig::C1B_bkg}. The spectrum covers an energy range of 1.5--200 keVee and is processed using anti-Compton veto and bulk-surface event discrimination~\cite{cdex1b2018}. The bulk-surface event discrimination is performed using the rise-time method described in Ref.~\cite{cdex1b2018}. No efficiency correction is applied in this analysis, which introduces only a small deviation within the energy range of 1.5--200 keVee. We are developing more appropriate bulk-surface event discrimination methods and corresponding efficiency correction procedures specifically for events in the high-energy region.

\section{\label{sec4} CDEX-1B background model}
To set constraints on inelastic WIMP dark matter from the CDEX-1B experiment, we construct the background model of CDEX-1B. We first obtain simulated spectra for each background component in the CDEX-1B experiment using \texttt{Geant4}~\cite{Geant4}. Thereafter, we determine the intensities of each component by fitting with the maximum likelihood method combined with the Markov chain Monte Carlo (MCMC) algorithm~\cite{MCMC}. This process yields an accurate background model of CDEX-1B.

\subsection{\label{sec4_1} Background simulation}
The background simulation of CDEX-1B is conducted with \texttt{Geant4}~\cite{Geant4}. In the simulation, we establish a complete CDEX-1B detector model and simulate the decays of radionuclides within detector components. 

The structural configuration of the detector comprises several critical components: the Ge crystal with inactive $\mathrm{N^+}$ layer, the crystal support structure, the signal pin, the front-end electronics, the vacuum chamber, and the NaI(Tl) anti-Compton detector. 

The inactive $\mathrm{N^+}$ layer of the Ge crystal exists only on the $\mathrm{N^+}$-doped surfaces, specifically the lateral and top sides. It consists of an outermost totally dead layer, where no charge can be collected, followed by an inner transition layer with partial charge collection~\cite{inactiveLayer,MJL:deadlayer}. According to Ref.~\cite{MJL:deadlayer}, the total inactive-layer thickness on the $\mathrm{N^+}$ surfaces of this crystal is 0.88 mm, including a 0.25 mm outer totally dead layer. The bottom surface, corresponding to the $\mathrm{P^+}$ point-contact electrode side, does not have an inactive $\mathrm{N^+}$ layer. It is coated with a very thin ($\sim$ 0.1~$\mathrm{\mu}$m) layer of amorphous germanium or silicon~\cite{P-layer}. 

In the \texttt{Geant4} simulation, energy depositions are recorded only when they occur in the active volume or the transition layer. Energy deposited in the totally dead layer is discarded, as it does not contribute to the observable signal. Events with any energy deposition in the transition layer are classified as surface events, whereas events depositing energy exclusively in the active volume are treated as bulk events. This procedure provides an effective description of the bulk-surface event discrimination without the need for explicit pulse-shape modeling.

We account for the following background sources in the simulation: cosmogenic radionuclides in the Ge crystal, radionuclides in surrounding structural materials, and radon progeny located on the outer surface of the vacuum chamber. All simulated radionuclides and their corresponding detector components are listed in Table~\ref{tab:bkg_component}. The ``Additional $^{210}$Pb'' in Table~\ref{tab:bkg_component} refers to the additional $^{210}$Pb contained in the lead materials inside the detector vacuum chamber, independent of the ``U Series'' and the ``Rn Progeny.''

\begin{table}[htbp]
  \caption{\label{tab:bkg_component}
  Simulated radionuclides and their corresponding detector components.}
  \renewcommand\arraystretch{1.5}
  \begin{ruledtabular}
  \begin{tabular}{cc}
            \textbf{Radionuclide} & \textbf{Component}\\
            \hline
            \multirow{1}{*}{\(^3\)H} & Crystal\\
            \hline
            \multirow{1}{*}{\(^{49}\)V} & Crystal\\
            \hline
            \multirow{1}{*}{\(^{55}\)Fe} & Crystal\\
            \hline
            \multirow{2}{*}{\(^{57}\)Co} & Crystal\\
                                         & Crystal Support Structure\\
            \hline
            \multirow{1}{*}{\(^{65}\)Zn} & Crystal\\
            \hline
            \multirow{1}{*}{\(^{68}\)Ge} & Crystal\\
            \hline
            \multirow{1}{*}{\(^{68}\)Ga} & Crystal\\
            \hline
            \multirow{1}{*}{\(^{73}\)As} & Crystal\\
            \hline
            \multirow{3}{*}{\(^{40}\)K} & Signal Pin\\
                                        & Signal Pin Support Structure\\
                                        & Front-End Electronics\\
            \hline
            \multirow{4}{*}{Additional \(^{210}\)Pb} & Crystal Support Structure\\
                                          & Signal Pin\\
                                          & Signal Pin Support Structure\\
                                          & Front-End Electronics\\
            \hline
            \multirow{4}{*}{Th Series} & Crystal Support Structure\\
                                       & Signal Pin\\
                                       & Signal Pin Support Structure\\
                                       & Front-End Electronics\\
            \hline
            \multirow{4}{*}{U Series} & Crystal Support Structure\\
                                      & Signal Pin\\
                                      & Signal Pin Support Structure\\
                                      & Front-End Electronics\\
            \hline
            \multirow{2}{*}{Ac Series} & Signal Pin Support Structure\\
                                       & Front-End Electronics\\
            \hline
            \multirow{2}{*}{Rn Progeny} & \multirow{2}{*}{\makecell{Outer Surface of\\ the Vacuum Chamber}}\\\\
  \end{tabular}
  \end{ruledtabular}
  \end{table}

We simulate at least $10^8$ decay events per background component to ensure statistical validity. The simulation also includes the anti-Compton veto of the NaI(Tl) anti-Compton detector~\cite{cdex1b2018}, bulk-surface event discrimination~\cite{MJL:deadlayer}, and the energy resolution of the CDEX-1B detector. 

By employing the background simulation, we derive the spectra for the background components of the CDEX-1B experiment.

\subsection{\label{sec4_2} Background fitting}

After obtaining the simulated spectra for each background component in CDEX-1B, we fit the experimental spectrum using the simulated spectra to determine their intensities.

We first use the optimizer in the \texttt{scipy.optimize} module~\cite{SciPy} to quickly obtain an approximate maximum-likelihood estimate. In practice, the optimizer minimizes the negative log-likelihood, which is mathematically equivalent to maximizing the likelihood. This estimate provides a reasonable starting point for the subsequent MCMC refinement.

The final optimal parameters are then obtained via MCMC sampling~\cite{MCMC} using the \texttt{emcee} package~\cite{emcee}. To efficiently reach the global maximum of the likelihood, the MCMC requires initial parameters that are close to the optimum, which are provided by the preceding \texttt{scipy.optimize} fit. Starting from these values, the MCMC algorithm explores the parameter space and converges to the global maximum.

To eliminate the influence of discrepancies in peak morphology between the experimental and simulated spectra on the fitting procedure, we merge each peak region defined by the identified peaks in the experimental spectrum into a single composite bins during spectral fitting.

\begin{figure*}[htbp]
  \includegraphics[width=\linewidth]{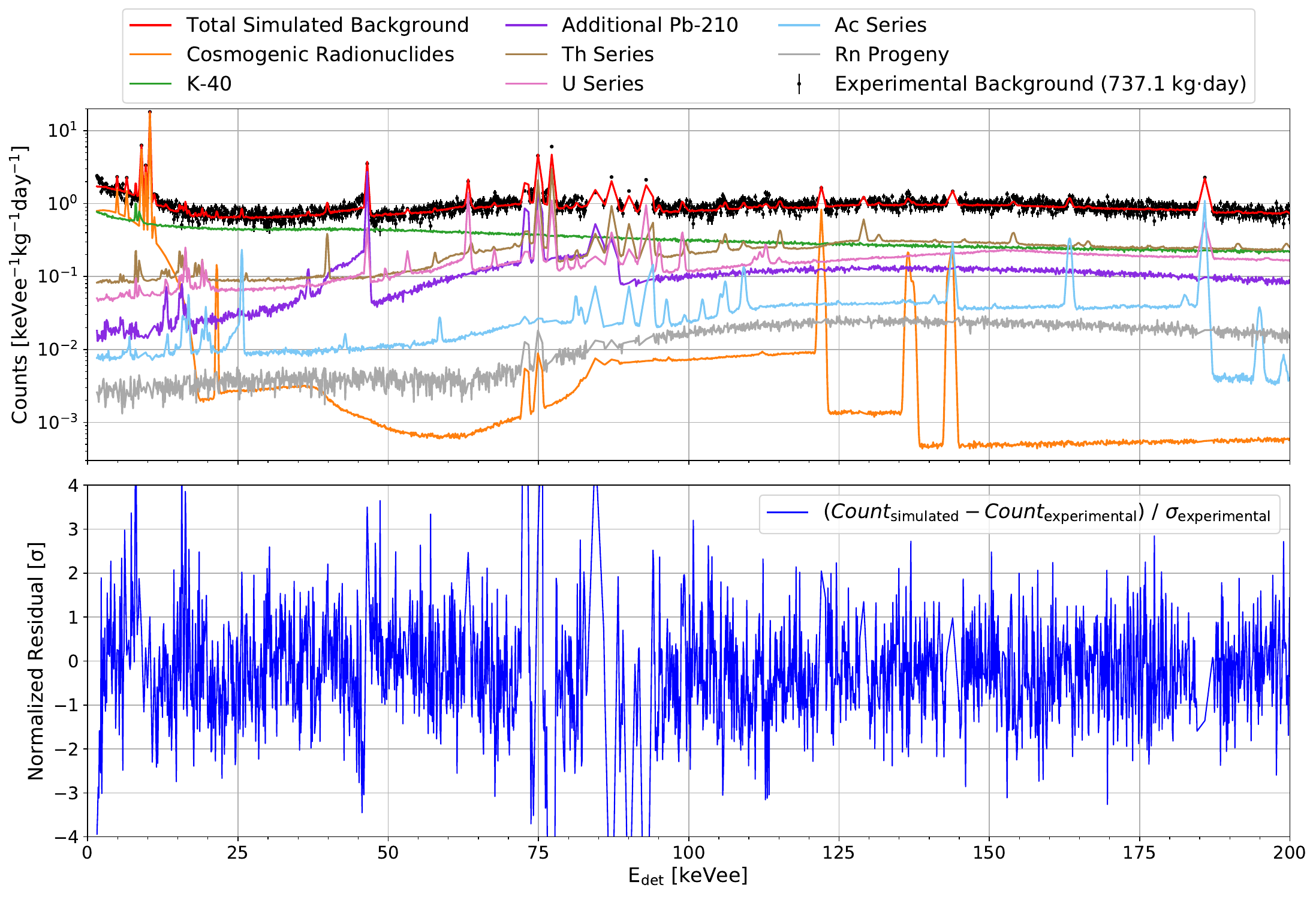}
  \caption{
  Background model of the CDEX-1B experiment. The upper panel displays the experimental background spectrum (black points with error bars) alongside the best-fit simulated spectrum (red line) and its constituent components (colored lines). The ``Additional Pb-210'' refers to the additional $^{210}$Pb contained in the lead materials inside the detector vacuum chamber, independent of the ``U Series'' and the ``Rn Progeny.'' The lower panel quantifies the agreement through normalized residuals $(Count_{\mathrm{simulated,\ i}} - Count_{\mathrm{experimental,\ i}})/\sigma_{\mathrm{experimental,\ i}}$, where $Count_{\mathrm{simulated,\ i}}$ and $Count_{\mathrm{experimental,\ i}}$ are simulated and experimental counts in bin i, respectively, and $\sigma{_\mathrm{experimental,\ i}}$ is the experimental error of bin i. Each peak region defined by the identified peaks in the experimental spectrum is merged into a single composite bin. Consequently, the peaks in the simulated spectra within these regions exhibit a triangular rather than a Gaussian profile in the upper panel.
  }
  \label{fig::fit_res}
  \end{figure*}

By combining the initial \texttt{scipy.optimize} fit and the subsequent MCMC refinement, we obtain the final optimal fit. This establishes an accurate background model for the CDEX-1B experiment. As shown in Fig.~\ref{fig::fit_res}, the background model reasonably reproduces the experimental spectrum. We note that the residuals between the simulated and experimental spectra in the 10--70 keV region are systematically negative, which affects the calculation of the final limits and results in conservative upper limits on the inelastic WIMP cross section.

\begin{figure}[htbp]
    \includegraphics[width=\linewidth]{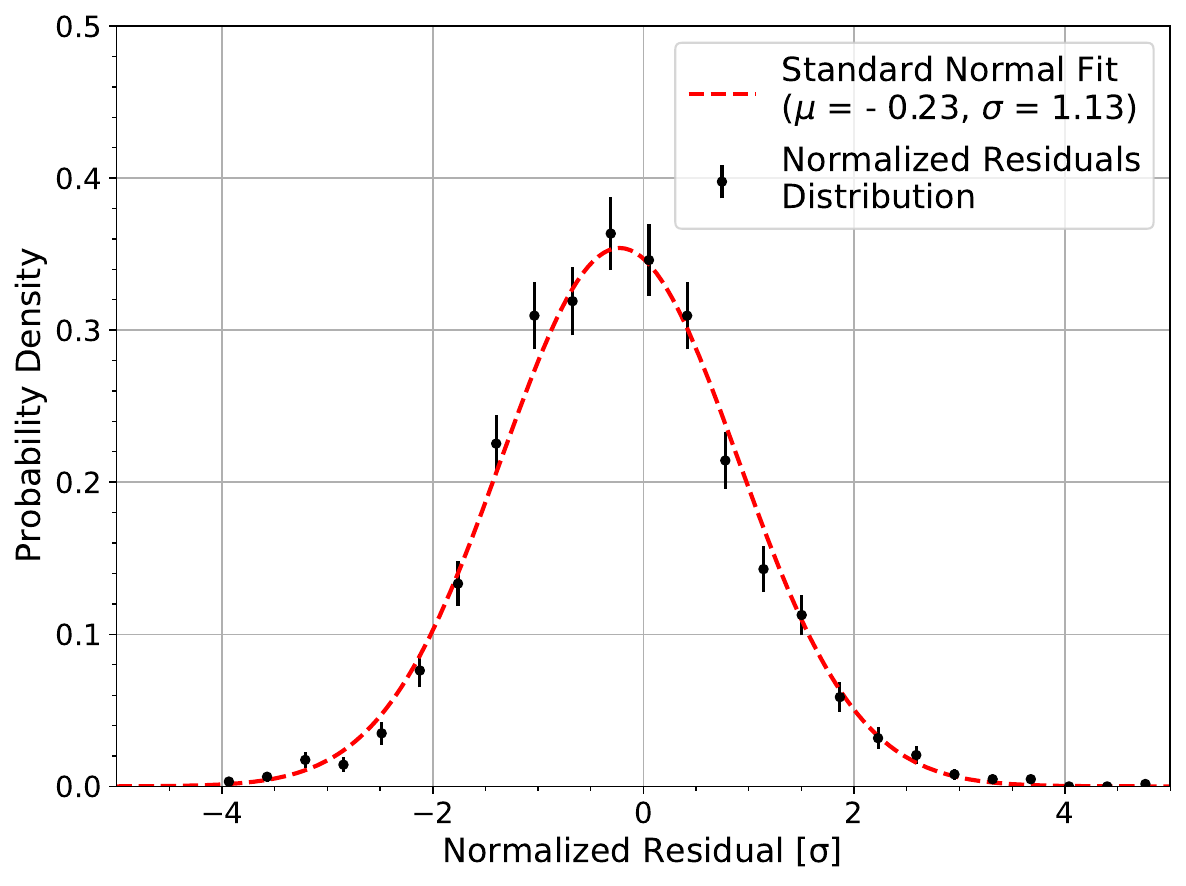}
    \caption{
      Distribution of the normalized residuals between the simulated and experimental counts for each energy bin (black points with error bars). A normal distribution fit to these residuals (red dashed line) yields a mean $\mu = -0.23$ and a standard deviation $\sigma = 1.13$. 
    }
    \label{fig::residual_dist}
    \end{figure}

  The distribution of normalized residuals between the simulated and experimental spectra, presented in Fig.~\ref{fig::residual_dist}, provides a quantitative evaluation of the fit quality. A normal distribution fit to the residuals yields a mean of approximately $-0.23$ and a standard deviation of about 1.13. The negative mean mainly reflects the systematic downward deviations in the 10--70 keV region. Aside from this effect, the residual distribution is reasonably close to a standard normal distribution, indicating that the overall fit quality is satisfactory.

\section{\label{sec5} Constraints on iDM}
Based on the expected spectra of iDM in the CDEX-1B detector and the background model of the CDEX-1B experiment, we calculate constraints on SI inelastic WIMPs from the CDEX-1B experiment.

\subsection{\label{sec5_1} Calculation method}
Following the methodology established during the construction of the background model, we employ a similar method based on the maximum likelihood fitting method and the MCMC algorithm~\cite{MCMC} to calculate constraints. 

First, we fit the experimental background spectrum of CDEX-1B using simulated spectra of each background component combined with the expected spectrum of inelastic WIMPs with given mass and splitting energy. The fitting utilizes the maximum likelihood method and the \texttt{scipy.optimize} module~\cite{SciPy}. The fit parameters include the intensities of background components and the WIMP signal, with initial values for the background intensities taken from the background model.

Subsequently, we apply the \texttt{emcee} implementation~\cite{emcee} of the MCMC algorithm to perform Bayesian parameter estimation, sampling the posterior distribution starting from the initial optimization results. 

In both fitting stages, we set the lower limit of the WIMP signal intensity to zero, meaning that negative intensities are excluded. Additionally, following the same methodology employed during the construction of the background model, we merge each peak region defined by the identified peaks in the experimental spectrum into a single composite bin during spectral fitting.  

By adopting the MCMC algorithm, we obtain the best-fit result and the posterior distribution of the WIMP signal intensity. Based on the posterior distribution of the WIMP signal intensity, we derive the posterior distribution of the corresponding $\sigma_{\mathrm{n}}$. The 90\% quantile of the posterior distribution of $\sigma_{\mathrm{n}}$ represents the 90\% confidence level (CL) upper limit (one sided) on it. 

\begin{figure}[htbp]
  \includegraphics[width=\linewidth]{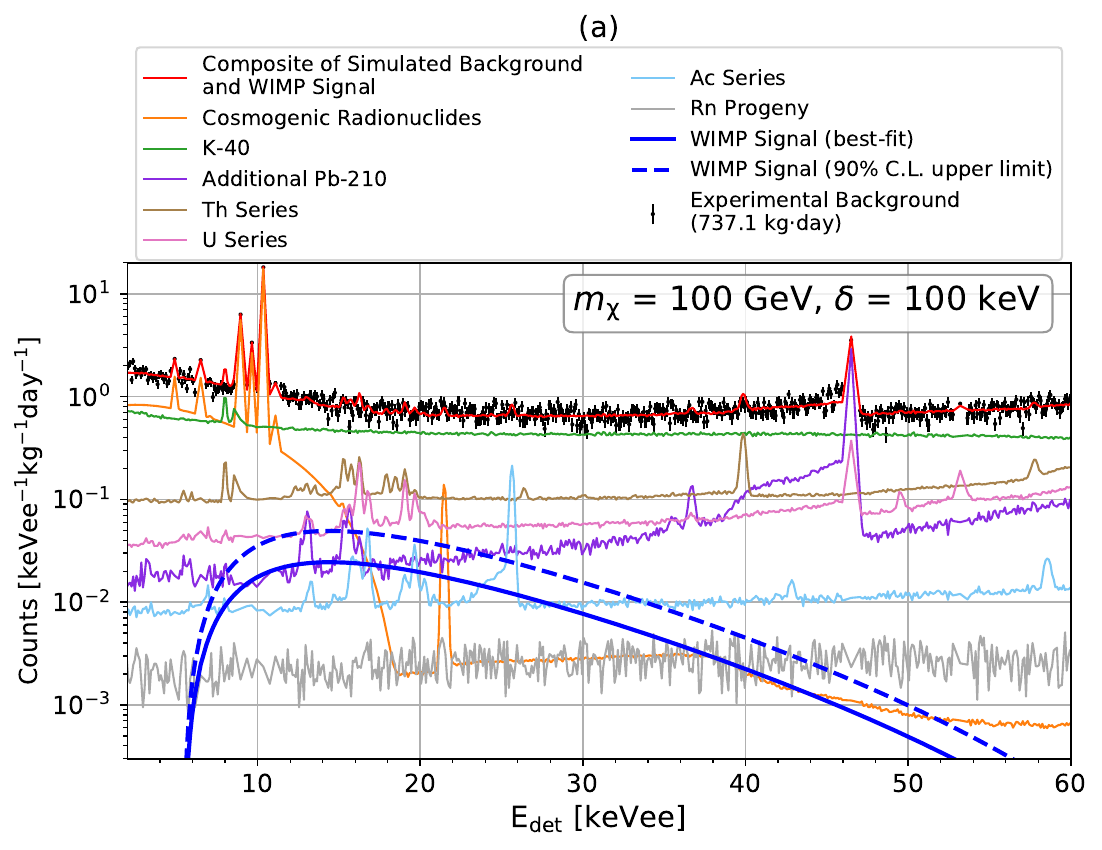}
  \includegraphics[width=\linewidth]{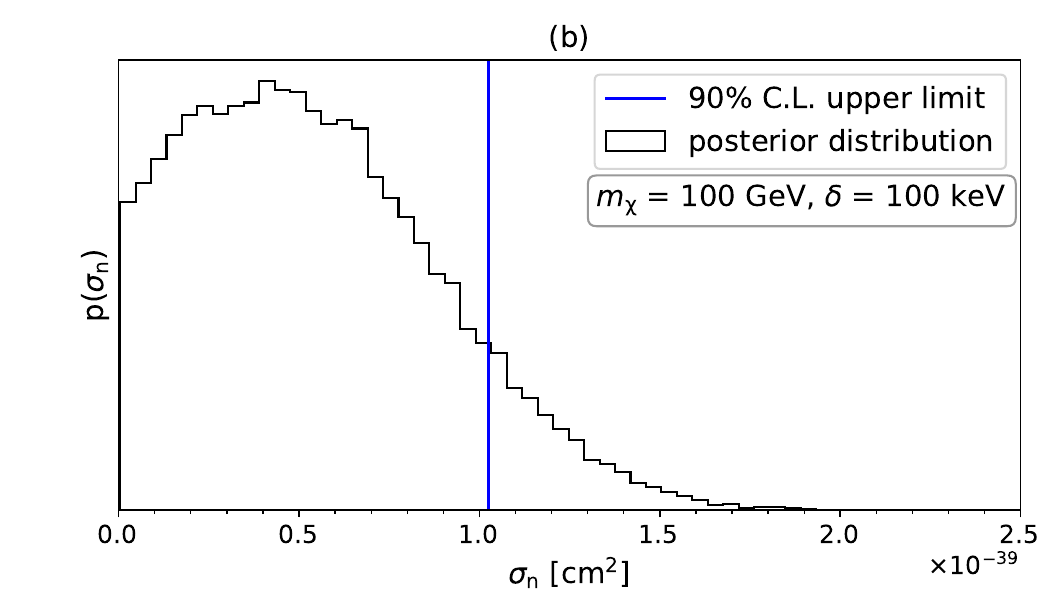}
  \caption{
  (a) Background and inelastic WIMP spectra corresponding to the best-fit result and the inelastic WIMP spectrum corresponding to the $90\%$ CL upper limits at $m_\chi = 100$ GeV, $\delta = 100$ keV. The black dots with error bars represent the experimental background spectrum. The red solid line represents the composite spectrum of the simulated background and the WIMP signal. The blue solid line represents the WIMP spectrum corresponding to the best-fit result, while the blue dashed line represents the WIMP spectrum corresponding to the 90\% CL upper limit. Solid lines in other colors indicate the simulated spectra of each background component. Each peak region defined by the identified peaks in the experimental spectrum is merged into a single composite bin. Consequently, the peaks in the simulated spectra within these regions exhibit a triangular rather than a Gaussian profile in the figure. (b) The posterior distribution of $\sigma_{\mathrm{n}}$ at $m_\chi = 100$ GeV, $\delta = 100$ keV. The blue vertical line indicates the 90\% CL upper limit on $\sigma_{\mathrm{n}}$.
  }
  \label{fig::fit_method}
  \end{figure}

Figure~\ref{fig::fit_method} illustrates the background and inelastic WIMP spectra corresponding to the best-fit result, the inelastic WIMP spectrum corresponding to the 90\% CL upper limit, and the posterior distribution of $\sigma_{\mathrm{n}}$ at $m_\chi = 100$ GeV, $\delta = 100$ keV. 

\subsection{\label{sec5_2} Results}

Using the above method, we calculate constraints on inelastic WIMPs for various $m_\chi$ and $\delta$ values. The low 1.5 keVee analysis threshold of CDEX-1B allows studies of the expanded parameter space with $\delta \ge 2$ keV and $m_\chi \ge 10$ GeV. Moreover, as can be seen from Eq.~(\ref{eq:vgm}), constrained by the Galactic escape velocity of WIMPs~\cite{SHM}, there exists an upper limit to the detectable $\delta$ for any given $m_\chi$ value. Additionally, as discussed in Sec.~\ref{sec4_2}, the systematic downward deviations of the background model relative to the experimental spectrum in the 10--70 keVee region lead to conservative limits on the inelastic WIMP cross section.

\begin{figure*}[htbp]
  \centering
  \begin{minipage}{0.48\textwidth}
    \centering
    \includegraphics[width=\linewidth]{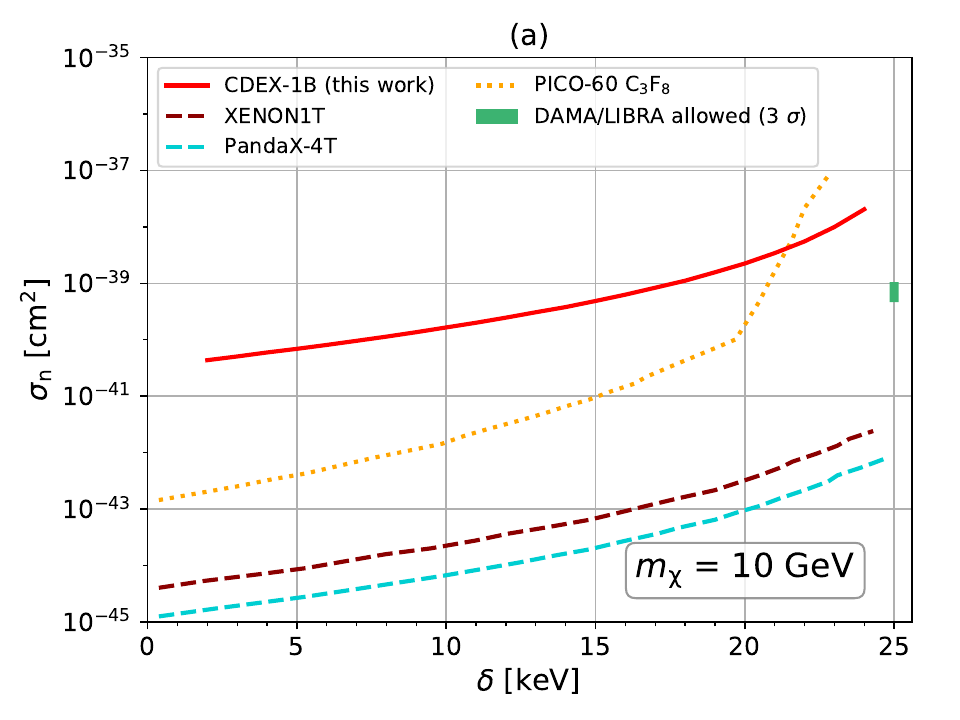}
  \end{minipage}%
  \hfill
  \begin{minipage}{0.48\textwidth}
    \centering
    \includegraphics[width=\linewidth]{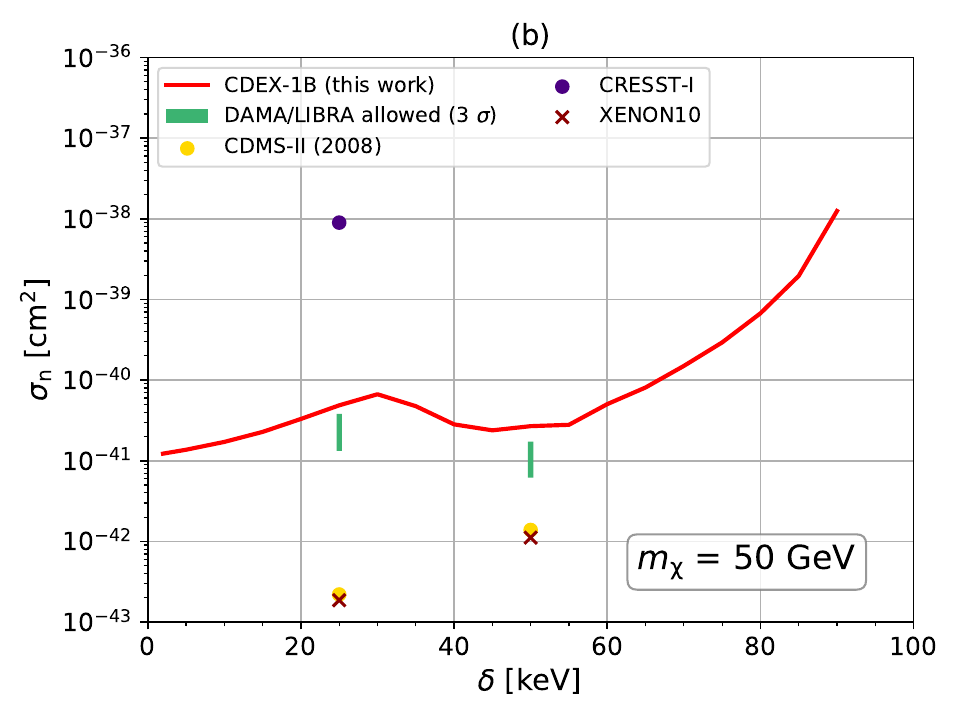}
  \end{minipage}
  
  \vspace{0.25cm}
  \begin{minipage}{0.48\textwidth}
    \centering
    \includegraphics[width=\linewidth]{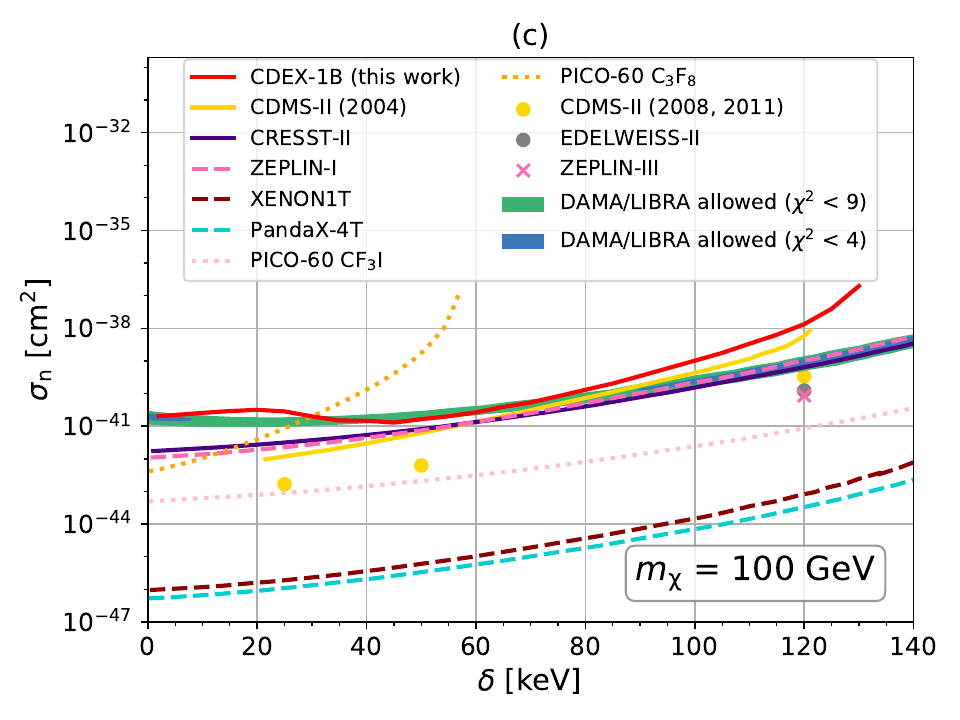}
  \end{minipage}%
  \hfill
  \begin{minipage}{0.48\textwidth}
    \centering
    \includegraphics[width=\linewidth]{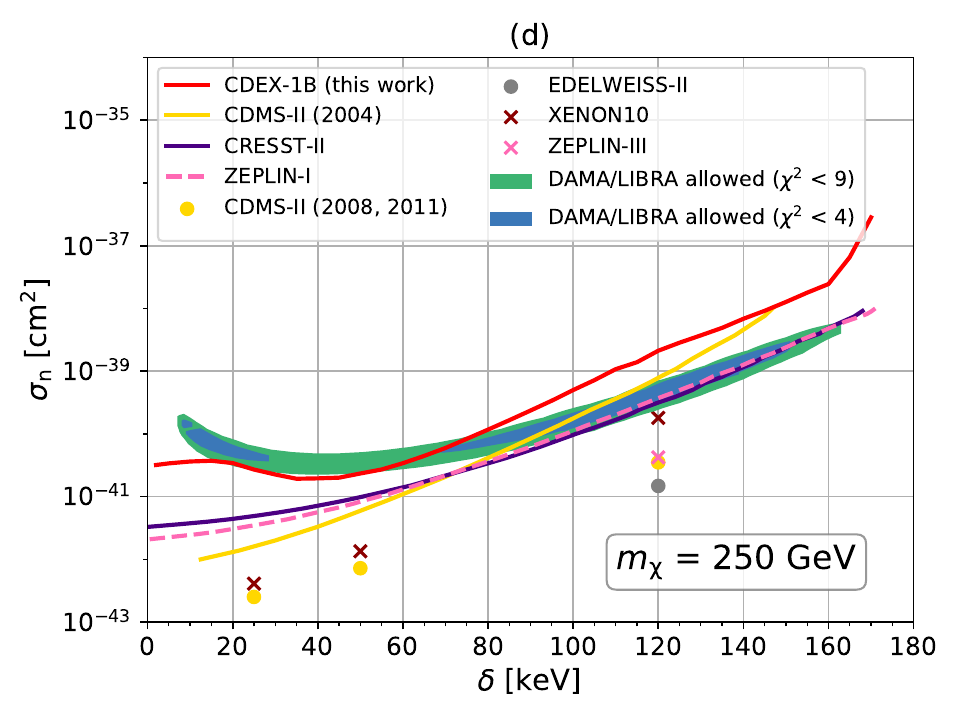}
  \end{minipage}
  
  \vspace{0.25cm}
  \begin{minipage}{0.48\textwidth}
    \centering
    \includegraphics[width=\linewidth]{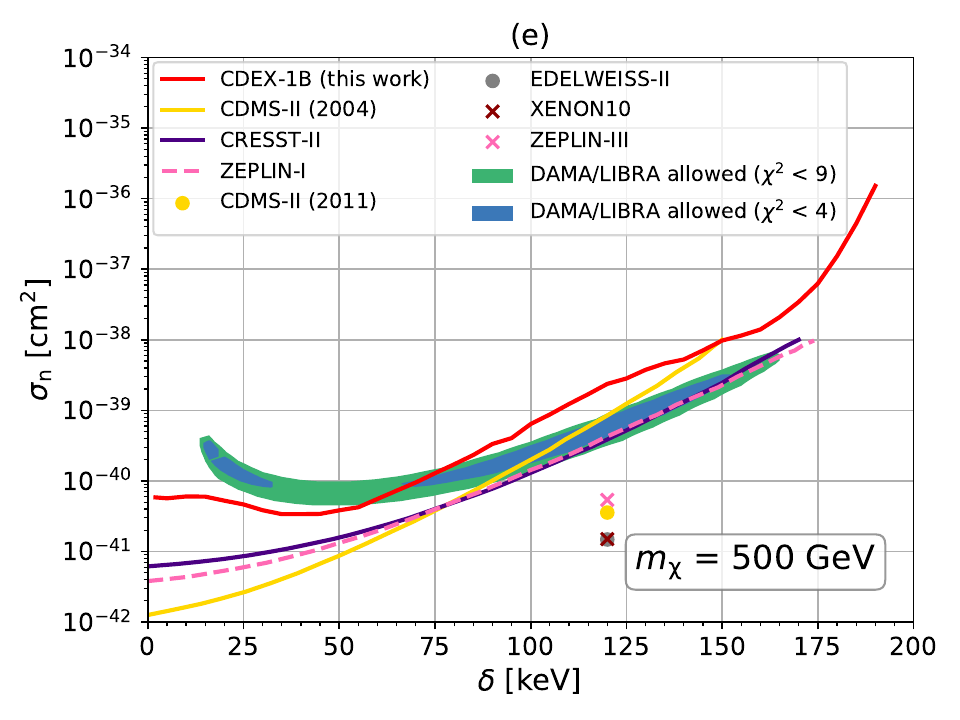}
  \end{minipage}%
  \hfill
  \begin{minipage}{0.48\textwidth}
    \centering
    \includegraphics[width=\linewidth]{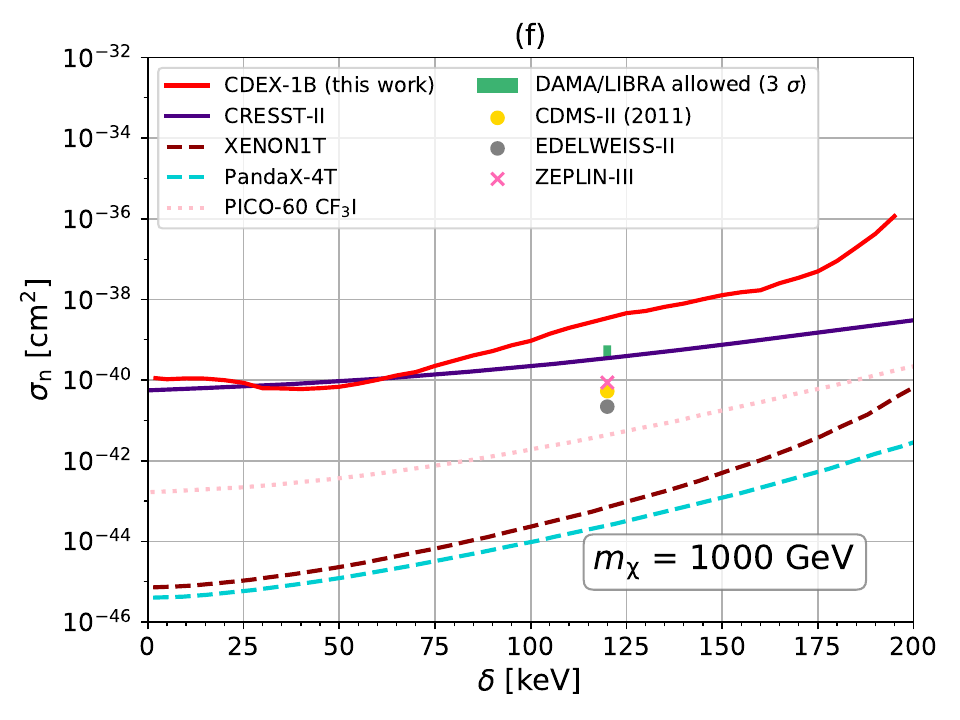}
  \end{minipage}  

  \caption{
    90\% CL upper limits on $\sigma_{\mathrm{n}}$ from CDEX-1B (red solid line) for $m_\chi = $ (a) 10 GeV, (b) 50 GeV, (c) 100 GeV, (d) 250 GeV, (e) 500 GeV, and (f) 1000 GeV. The 90\% CL upper limits from XENON1T (dark red dashed line)~\cite{XENON1T:2018,PICO}, PandaX-4T (cyan dashed line)~\cite{PandaX-4T:2021,PICO}, PICO-60 (orange dotted line for $\mathrm{C_3F_8}$ limits, pink dotted line for $\mathrm{CF_3I}$ limits)~\cite{PICO}, CDMS-II (gold solid line)~\cite{CDMS1,iDMstatus}, CRESST-II (indigo solid line)~\cite{cresst2,CRESST:2015,iDMstatus,InelasticFrontier}, and ZEPLIN-I (hot pink dashed line)~\cite{ZEPLIN1,iDMstatus}are also shown. Results from CDMS-II (gold circle marker)~\cite{CDMS-II:2008,CDMS-II:2011,DAMAandWIMP,EDELWEISS-II}, CRESST-I (indigo circle marker)~\cite{cresst1,DAMAandWIMP}, XENON10 (dark red cross marker)~\cite{XENON10:2007,XENON10:2009,DAMAandWIMP,EDELWEISS-II}, EDELWEISS-II (gray circle marker)~\cite{EDELWEISS-II}, and ZEPLIN-III (hot pink cross marker)~\cite{ZEPLIN-III:2010, EDELWEISS-II} that provide limits only at certain specific $\delta$ values for some $m_\chi$ values are indicated using markers, with each marker representing a 90\% CL upper limit. The regions allowed by DAMA/LIBRA at $\chi^2 < 9$ (green shaded) and $\chi^2 < 4$ (blue shaded)~\cite{DAMA1,DAMA2,iDMstatus} are represented as shaded areas, and the DAMA/LIBRA 3$\sigma$ allowed regions~\cite{DAMA:2008,DAMA2010,CDMS-II:2011,DAMAandWIMP,EDELWEISS-II} corresponding to $m_\chi$ values where only certain $\delta$ values have been calculated are indicated by green vertical line-shaped regions.
  }
  \label{fig::res_delta}
\end{figure*}

Figure~\ref{fig::res_delta} shows the 90\% CL upper limits on $\sigma_{\mathrm{n}}$ from the CDEX-1B experiment for $m_\chi = 10$, 50, 100, 250, 500, and 1000 GeV. The 90\% CL upper limits from XENON10~\cite{XENON10:2007,XENON10:2009,DAMAandWIMP,EDELWEISS-II}, XENON1T~\cite{XENON1T:2018,PICO}, PandaX-4T~\cite{PandaX-4T:2021,PICO}, PICO-60~\cite{PICO}, CDMS-II~\cite{CDMS1,CDMS-II:2008,CDMS-II:2011,iDMstatus,DAMAandWIMP,EDELWEISS-II}, CRESST-I~\cite{cresst1,DAMAandWIMP}, CRESST-II~\cite{cresst2,CRESST:2015,iDMstatus,InelasticFrontier}, ZEPLIN-I~\cite{ZEPLIN1,iDMstatus}, ZEPLIN-III~\cite{ZEPLIN-III:2010,EDELWEISS-II}, and EDELWEISS-II~\cite{EDELWEISS-II} are also shown. Results from CDMS-II~\cite{CDMS-II:2008,CDMS-II:2011,DAMAandWIMP,EDELWEISS-II}, CRESST-I~\cite{cresst1,DAMAandWIMP}, XENON10~\cite{XENON10:2007,XENON10:2009,DAMAandWIMP,EDELWEISS-II}, EDELWEISS-II~\cite{EDELWEISS-II}, and ZEPLIN-III~\cite{ZEPLIN-III:2010, EDELWEISS-II}, which provide limits only at certain specific $\delta$ values for some $m_\chi$ values, are indicated using markers, with each marker representing a 90\% CL upper limit. The DAMA/LIBRA allowed regions at $\chi^2 < 9$ (green shaded) and $\chi^2 < 4$ (blue shaded)~\cite{DAMA1,DAMA2,iDMstatus} are represented as shaded areas, while the DAMA/LIBRA 3$\sigma$ allowed regions~\cite{DAMA:2008,DAMA2010,CDMS-II:2011,DAMAandWIMP,EDELWEISS-II} corresponding to $m_\chi$ values where only certain $\delta$ values have been calculated are indicated by vertical line-shaped regions. At $m_\chi = 250$ GeV, the CDEX-1B upper limits fully exclude the $\chi^2 < 4$ allowed region from DAMA/LIBRA for $\delta < 30$ keV, and at $m_\chi = 500$ GeV, they fully exclude the $\chi^2 < 9$ allowed region for $\delta < 50$ keV.

\section{\label{sec6} Conclusions and discussions}
In this work, we present the SI inelastic WIMP dark matter search results from the CDEX-1B experiment~\cite{cdex1b2018}. By establishing an accurate background model of the CDEX-1B experiment, we calculate constraints on inelastic WIMPs using the CDEX-1B dataset with an exposure of 737.1 kg$\cdot$day~\cite{cdex1b2018}. The $90\%$ CL upper limits on the SI WIMP-nucleon cross section $\sigma_{\mathrm{n}}$ from CDEX-1B fully exclude the $\chi^2 < 4$ allowed regions for $\delta <$ 30 keV at $m_\chi =$ 250 GeV and the $\chi^2 < 9$ allowed region for $\delta <$ 50 keV at $m_\chi$ = 500 GeV from DAMA/LIBRA~\cite{DAMA1,DAMA2,iDMstatus}.

A method is developed in this paper to establish the CDEX-1B background model and to place constraints on iDM. The analysis procedures can be adopted to probe various variants of iDM models~\cite{MiDM,EFTiDM,iDDM,iBDM1,iBDM2}. We note that the systematic downward deviations of the background model relative to the experimental spectrum in the 10--70 keVee region lead to conservative limits on inelastic WIMPs. In future analyses, we plan to further optimize the background fit to reduce these deviations, which is expected to provide more precise constraints on inelastic WIMP interactions.

Future studies could explore the detection of decay products from excited iDM~\cite{iBDM2,iDMdecay}. Additionally, since iDM is more sensitive to the velocity distribution of dark matter compared to elastic dark matter, better results may be achieved by using the modulation method for analysis~\cite{SI_iDM,cdex1b_am,DAMAmodule}.

The next-generation of the CDEX experiment, CDEX-50~\cite{CDEX50pre}, is being prepared. In the CDEX-50 experiment, an upgraded detector array consisting of 50 HPGe detectors with a target mass of 50 kg will be deployed. The CDEX-50 experiment has a target exposure of 150 kg$\cdot$year, and its background level will be reduced to approximately $2 \times 10^{-4}$ $\mathrm{counts \cdot kg^{-1} \cdot keVee^{-1}  \cdot day^{-1}}$ at 20 keVee~\cite{CDEX50pre}, which is $5 \times 10^3$ times lower than that of CDEX-1B. Therefore, the results of the CDEX-50 experiment are expected to improve limits on $\sigma_{\mathrm{n}}$ by 4 orders of magnitude compared to our current results.

\acknowledgments
This work was supported by the National Key Research and Development Program of China (Grants No. 2023YFA1607101 and No. 2022YFA1605000) and the National Natural Science Foundation of China (Grants No. 12322511 and No. 12175112). We acknowledge the Center of High Performance Computing, Tsinghua University, for providing the facility support. We would like to thank CJPL and its staff for hosting and supporting the CDEX project. CJPL is jointly operated by Tsinghua University and Yalong River Hydropower Development Company.

\FloatBarrier
\bibliography{iDM}

\end{document}